\documentclass[conference]{IEEEtran}
\IEEEoverridecommandlockouts
\usepackage{cite}
\usepackage{amsmath,amssymb,amsfonts}
\usepackage{algorithmic}
\usepackage{graphicx}
\usepackage{textcomp}
\usepackage{xcolor}
\usepackage{mathtools}
\usepackage{multirow}

\usepackage{textcomp}

\newcommand{\STAB}[1]{\begin{tabular}{@{}c@{}}#1\end{tabular}}

\def\BibTeX{{\rm B\kern-.05em{\sc i\kern-.025em b}\kern-.08em
	T\kern-.1667em\lower.7ex\hbox{E}\kern-.125emX}}
\begin{document}

\title{Wake-Cough: cough spotting and cougher identification for personalised long-term cough monitoring \\
	\thanks{This project was funded by the South African Medical Research Council (SAMRC) through its Division of Research Capacity Development under the SAMRC Intramural Postdoctoral programme, the South African National Treasury, as well as an EDCTP2 programme supported by the European Union (grant TMA2017CDF-1885, grant SF1401, OPTIMAL DIAGNOSIS; grant RIA2020I-3305, CAGE-TB) and the National Institute of Allergy and Infection Diseases of the National Institutes of Health (U01AI152087). We also thank the South African Centre for High Performance Computing (CHPC) for providing computational resources on their Lengau cluster for this research, and gratefully acknowledge the support of Telkom, South Africa.}
}

\author{Madhurananda Pahar$^{1}$, Marisa Klopper$^{2}$, Byron Reeve$^{2}$, Rob Warren$^{2}$, Grant Theron$^{2}$, \\ Andreas Diacon$^{3}$ and Thomas Niesler$^{1}$%
	
	\vspace{10pt}
	\\

	\textit{$^1$Department of Electrical and Electronic Engineering, Stellenbosch University, South Africa}\\
	\textit{$^2$Division of Molecular Biology and Human Genetics,} \textit{Stellenbosch University, South Africa}\\
	\textit{$^3$TASK Applied Science, Cape Town, South Africa}
	
	\vspace{8pt}
	\\
	Email: \{mpahar, marisat, byronreeve, rw1, gtheron, ahd, trn\}@sun.ac.za
	
}

\maketitle

\begin{abstract}
	We present `wake-cough', an application of \mbox{wake-word} spotting to coughs
	using a Resnet50 and the identification of coughers using i-vectors, for the purpose of a \mbox{long-term}, personalised cough monitoring system. 
	Coughs, recorded in a quiet (73$\pm$5 dB) and noisy (34$\pm$17 dB) environment, were used to extract i-vectors, x-vectors and \mbox{d-vectors}, used as features to the classifiers.
	The system achieves 90.02\% accuracy when using an MLP to discriminate between 51 coughers using 2-sec long cough segments in the noisy environment. 
	When discriminating between 5 and 14 coughers using longer (100 sec) segments in the quiet environment, this accuracy improves to 99.78\% and 98.39\% respectively. 
	Unlike speech, i-vectors outperform x-vectors and d-vectors in identifying coughers. 
	These coughs were added as an extra class to the Google Speech Commands dataset and features were extracted by preserving the \mbox{end-to-end} \mbox{time-domain} information in a trigger phrase. 
	The highest accuracy of 88.58\% is achieved in spotting coughs among 35 other trigger phrases using a Resnet50. 
	Thus, \mbox{wake-cough} represents a personalised, \mbox{non-intrusive} cough monitoring system, which is power-efficient as on-device \mbox{wake-word} detection can keep a smartphone-based monitoring device mostly dormant.  
	This makes \mbox{wake-cough} extremely attractive in multi-bed ward environments to monitor patients' long-term recovery from lung ailments such as tuberculosis (TB) and COVID-19.
\end{abstract}

\section{INTRODUCTION}

Wake-words are used as trigger phrases which enable keyword spotting systems to initiate certain tasks such as speech recognition 
by continuously listening for specific keywords using low computational power \cite{schalkwyk2010your}. 
This is the first important step between the user and the processing units on either the device or the cloud server \cite{wu2018monophone} and both the near and far field wake-word detection requires to be highly sensitive in both quiet and noisy environments for better performance \cite{gao2020towards}. 
For example, some widely-used trigger phrases for voice assistants on smart devices are: Google's `OK Google', Apple's `Hey Siri', Amazon's `Alexa' and Microsoft's `Hey Cortana' 
\cite{sainath2015convolutional}.
These algorithms are highly sensitive in both quiet and noisy environments \cite{gao2020towards},
making them extremely useful in hands-free situations like driving \cite{chen2014small}.
Coughing is the forceful expulsion of air to clear the airway and a common symptom of 
respiratory diseases, 
such as tuberculosis (TB) \cite{botha2018detection}, asthma \cite{al2013signal}, pertussis~\cite{pramono2016cough} and \mbox{COVID-19} \cite{pahar2020covid, pahar2022covid}, which can be identified using machine learning classifiers.
To successfully implement cough as a personalised wake-word in commercial smartphones, it is necessary to accurately identify the cougher \cite{ge2017deep} in both noisy and quiet environments and the cough among various other commonly used trigger phrases \cite{kepuska2009novel}.

Vocal audio such as speech can be identified using
\mbox{i-vectors}, 
which present a low-dimensional speaker and channel-dependant space using factor analysis
proposing a speaker representation system for speaker identification 
\cite{senior2014improving}.
The performance can be improved by using x-vectors \cite{snyder2018x} and d-vectors \cite{wan2018generalized}, which
use the data augmentation and DNN based embeddings to map speaker embeddings.


Coughers have been identified using x-vectors on natural coughs in an open world environment for 8 male and 8 female subjects after implementing data augmentation to address the effect of background noise \cite{whitehill2020whosecough}
and using \mbox{d-vectors} on
forced coughs 
\cite{zhang2017speaker}. 
Here, we identify both natural and forced coughs among other trigger phrases in the Google Speech commands dataset \cite{warden2018speech} while also identifying the coughers in noisy and quiet environments using \mbox{i-vectors}, \mbox{x-vectors} and d-vectors. 
To accurately monitor the \mbox{long-term} cough rates, for example in a multi-bed ward, automatic detection of coughs among other environmental noises and classification of coughers while consuming less power and preserving privacy is extremely important. 
By detecting coughs among other wake-words and classifying coughers using i-vectors, wake-cough represents a personalised long-term cough monitoring system. 
This system is also \mbox{power-efficient} as specialised algorithms work on the device without needing any cloud service.




\section{Dataset Preparation}

\begin{figure}[!h]
\centerline{\includegraphics[width=0.5\textwidth]{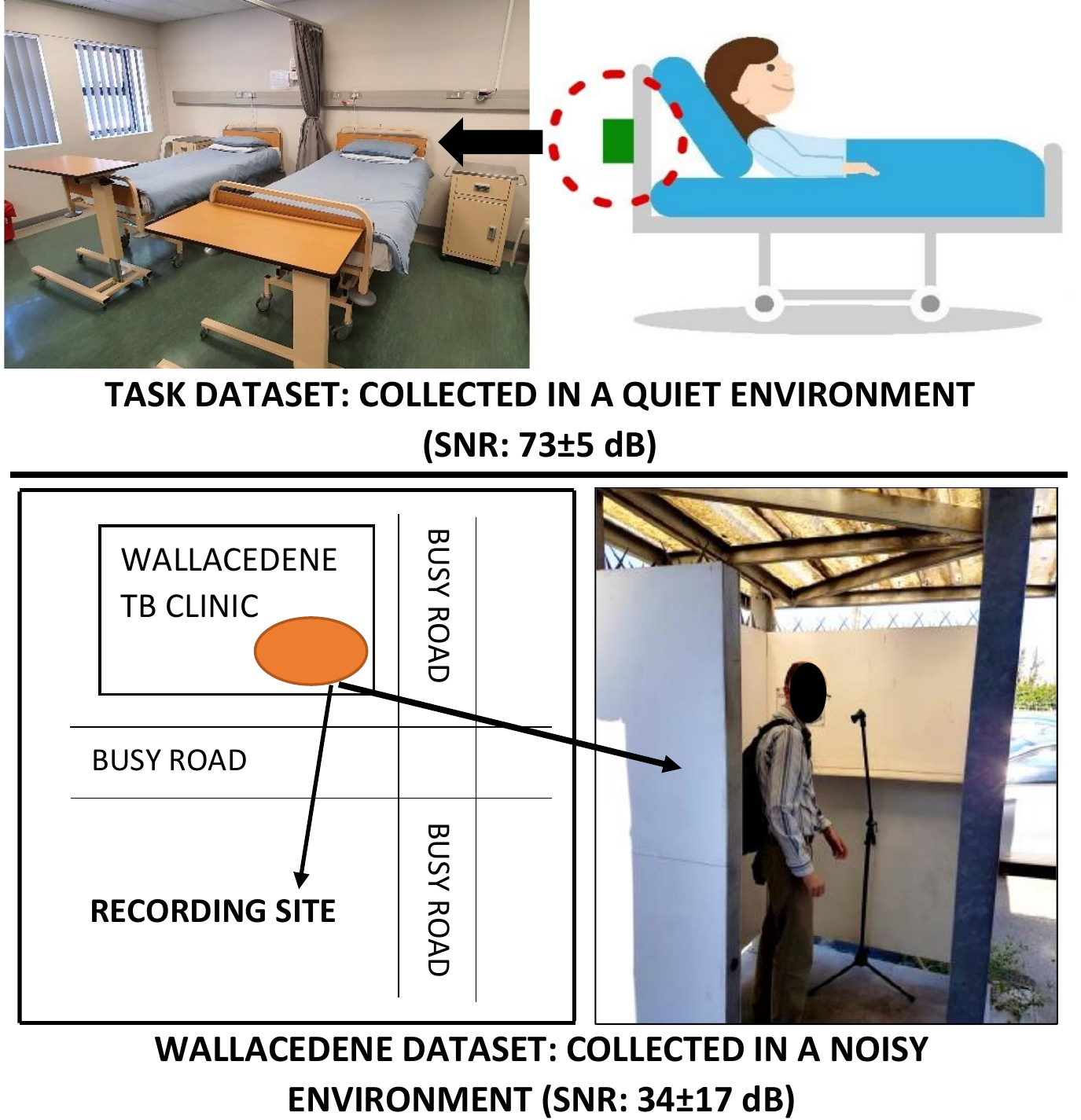}}
\caption{\textbf{Data collection process for cougher identification:} The TASK dataset, containing only coughs, was collected in a quiet environment. The Wallacedene dataset, containing both cough and speech (counting from 1 to 10), was collected in a noisy environment. }
\label{fig:data-collection}
\end{figure}

For the cougher identification task, two datasets which will be referred to as TASK and Wallacedene (Table \ref{table:dataset-summary}), were both manually annotated using ELAN \cite{wittenburg2006elan}. 
The TASK dataset, which contains natural coughs, was collected at a TB research hospital in Cape Town, South Africa (TASK clinical trial centre). 
This research hospital accommodates up to 24 patients in six 4-bed wards \cite{pahar2021deep, pahar2022automatic}. 
A plastic enclosure, attached to the bed-frames, holds a Samsung Galaxy J4 smartphone connected to a BOYA BY-MM1 cardioid microphone 
(Figure \ref{fig:data-collection}) and
the distance between the cougher and the microphone was between 30 and 150 cm.
The dataset includes 6000 cough events, sampled at 22.05 kHz and collected from 14 adult male patients over a 6 month period, totalling 3.16 hours of cough audio with an average SNR of 73$\pm$5 dB.
No other information of the patients was collected due to ethical constraints. 
Wallacedene dataset was collected inside an outdoor booth next to a busy primary health clinic in Wallacedene, near Cape Town, South Africa representing a real-world environment where a typical TB test would likely to be deployed \cite{pahar2021tb}  (Figure \ref{fig:data-collection}). 
Patients were asked to count from 1 to 10, then cough, take a few deep breaths, and cough again, thus producing a bout of forced coughs. 
These counts were used as speech to provide a baseline to compare the performance of cougher identification in Table \ref{table:cougher-ident}. 
The audio, sampled at 44.1 kHz, was recorded using
a R{\O}DE M3 condenser microphone 
from 38 males and 13 females, keeping a 10 to 15 cm gap between the microphone and the patients. 
Environmental noise was present in both cough and speech recordings, which had an average SNR of 34 dB and 33 dB respectively with a standard deviation of 17 dB (Table \ref{table:dataset-summary}).

Table \ref{table:dataset-summary} shows that the TASK dataset is less-noisy and contains much longer cough audio for each subject, whereas the Wallacedene dataset is noisier but contains both cough and speech audio from a larger number of subjects. 
All audio recordings were downsampled to 16 kHz, as required for the Kaldi ASR system \cite{povey2011kaldi}.

\begin{table}[h]

\small

\setlength{\tabcolsep}{5pt} 

\caption{\textbf{Data used in cougher \& speaker identification}: The TASK dataset is less-noisy than the Wallacedene dataset.  } 
\centering 
\begin{center}
	\begin{tabular}{ l | c c c c }
		\hline
		\hline
		\textbf{Dataset} & \textbf{Subjects} & \textbf{Events} & \textbf{Avg SNR} & \textbf{Avg Length}\\
		\hline
		\hline
		\multicolumn{5}{c}{\textbf{\textit{Cougher identification}}}\\
		\hline
		\hline
		TASK & 14  &  6000 &  73$\pm$5 dB & 1.87$\pm$0.2 sec \\
		
		\hline
		Wallacedene & 51  & 1358  & 34$\pm$17 dB  &  0.77$\pm$0.1 sec \\
		
		\hline
		\hline
		
		\multicolumn{5}{c}{\textbf{\textit{Speaker identification}}}\\
		\hline
		\hline
		
		Wallacedene & 51  & 510  & 33$\pm$17 dB  &  0.99$\pm$0.2 sec \\
		\hline
		\hline
	\end{tabular}
\end{center}
\label{table:dataset-summary}
\end{table}


\vspace{-5pt}

For cough spotting, we randomly selected 3795 coughs from the TASK and Wallacedene datasets.
Each cough was normalised to a 1-sec duration by either trimming or padding with silence.
These `cough' events were added as an extra class to the 2nd version of Google Speech Commands dataset, which contains a total of 109,624 1-sec long  events, 
sampled at 16 kHz and belonging to 35 classes \cite{warden2018speech}.
These events were mixed with the background noises 
(Section 5.8 of \cite{warden2018speech}) 
with a randomly selected SNR between 73 and 34 dB (Table \ref{table:dataset-summary}). 
A subset of this dataset,
with only 42,341 events belonging to 10 classes, is also available for use as commands in IoT or robotics \cite{warden2018speech}. 
For spotting cough as a trigger phrase, we note these two datasets as SC-36 and SC-11, containing 36 and 11 classes respectively.




\section{Feature Extraction}

\begin{figure}[!h]
\centerline{\includegraphics[width=0.5\textwidth]{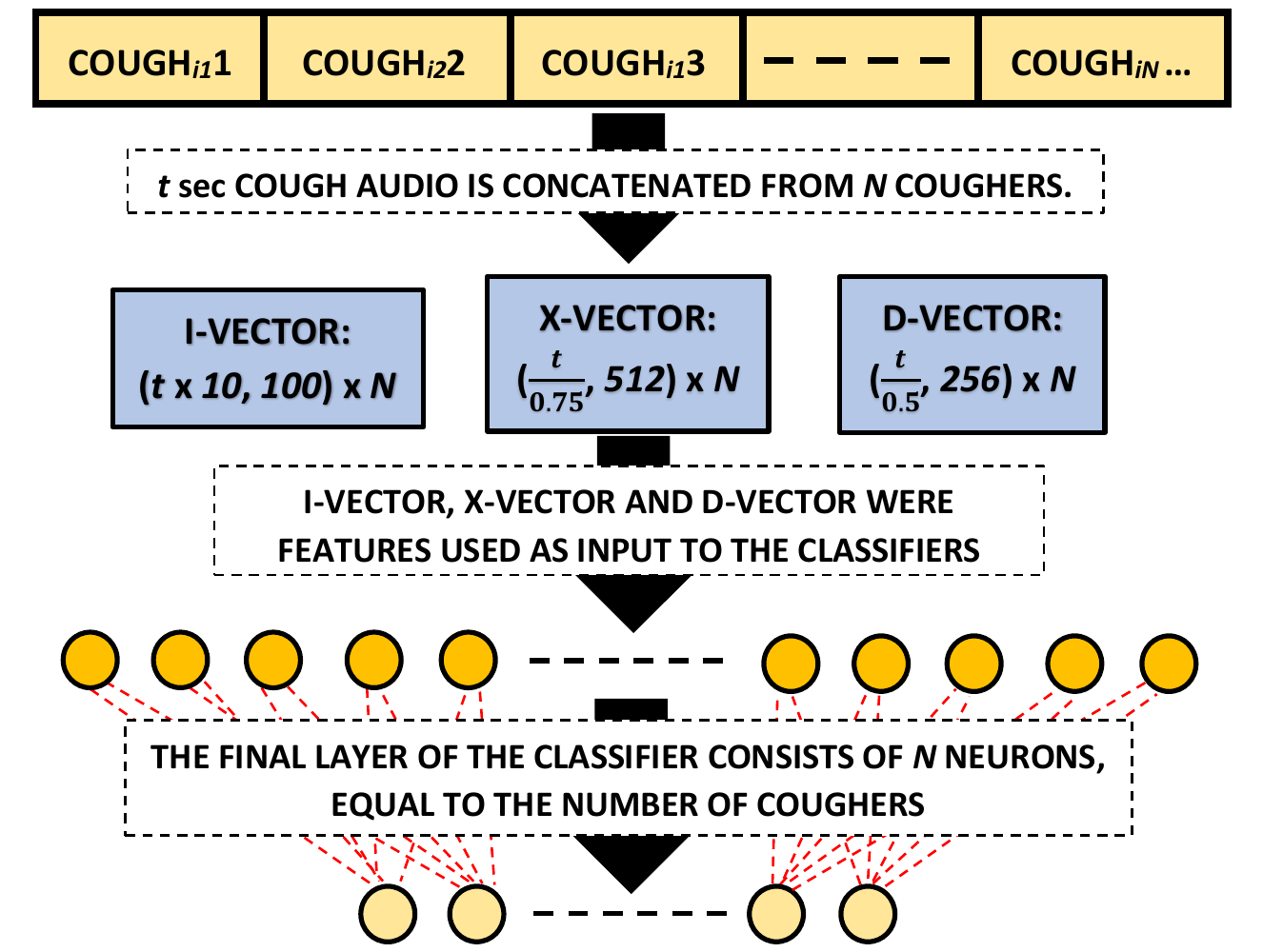}}
\caption{\textbf{Feature extraction for cougher identification:} $t$-sec long cough segments ($\text{COUGH}_{i1}$, $\text{COUGH}_{i2}$, $\text{COUGH}_{i3}$, $\ldots$, where, $1 \leq i \leq N$) from each cougher are concatenated as they appear in the audio recording for $N$ coughers. i-vectors, x-vectors and d-vectors are extracted from this $t \times N$-sec long audio and presented to the classifiers, which have $N$ neurons in the final layer to distinguish the cougher using a cross-validation scheme. }
\label{fig:feat-extract}
\end{figure}

For cougher identification, 
we have extracted x-vectors and i-vectors using extractors pre-trained on the under-resourced languages \cite{padhi2020}, which are spoken by the subjects in the TASK and Wallacedene datasets (Figure \ref{fig:feat-extract}). 
Audio segments that are $t$-sec long from each of $N$ coughers are concatenated by following the data preparation requirements of Kaldi ASR toolkit \cite{povey2011kaldi}. 
For each non-overlapping 0.1 sec audio, i-vectors are generated from each utterance ID, with a dimension of $(t \times 10, 100)$ for each cougher 
\cite{senior2014improving}.
Unique x-vectors are generated for each 1.5 sec of utterance with a 0.75 sec overlap, having a dimension of $(1, 512)$ \cite{snyder2018x}.
Thus for each $t$-sec long audio from each cougher, there are x-vectors of dimension $(\frac{t}{0.75}, 512)$. 
We have also extracted d-vectors using an extractor pre-trained on VCC 2018, VCTK, LibriSpeech, and CommonVoice English datasets and were generalized using the end-to-end loss function \cite{wan2018generalized}. 
Every $t$ sec cough is split into non-overlapping 0.5 sec segments, thus producing \mbox{d-vectors} of dimension $(\frac{t}{0.5}, 256)$ for every cougher and suggesting that the i-vectors have a higher dimensionality than x-vectors and d-vectors. 
The number of subjects ($N$) and the cough-time ($t$) were the hyperparameters in cougher identification task (Table \ref{table:class-hyper-parameter}). 
For speakers, we used all counts, 
having only $N$ as a hyperparameter. 
For the TASK and Wallacedene datasets, $N$ has been varied between 5 \& 14 and 5 \& 51 respectively
in steps of 5.

For spotting cough as a trigger phrase,
we have extracted STFT, ZCR and kurtosis
from overlapping frames ($\mathcal{F}$) of the audio, where the frame overlap is computed to ensure that the audio signal is always divided into exactly $\mathcal{S}$ frames, so that
the entire audio event is always captured within a fixed number of frames, allowing a fixed input dimension to be maintained while preserving the general overall temporal structure of the event.
Such fixed two-dimensional features are particularly useful for the training of DNN classifiers \cite{pahar2020covid}.
Table~\ref{table:feat-hyper-parameter} shows that in our experiments each audio signal is divided into between 70 and 150 frames, each between 512 and 4096 samples i.e. 32 msec and 256 msec long, thus varying
the spectral information extracted from each event in the \mbox{SC-11} and SC-36 datasets. 


\begin{table}[h]
\footnotesize
\setlength{\tabcolsep}{3pt} 
\caption{\textbf{Feature extraction hyperparameters}. Table \ref{table:cougher-ident} and \ref{table:cough-KWS} show classification results for these hyperparameters. } 
\centering 
\begin{center}
	\begin{tabular}{ c | c | c }
		\hline
		\hline
		\textbf{Hyperparameter} & \textbf{Description} & \textbf{Range} \\
		\hline
		\hline
		\multicolumn{3}{c}{\textbf{\textit{Cougher identification}}}\\
		\hline
		\hline
		Subject ($N$) & no. of coughers or speakers & 5 to 51 with step of 5 \\
		\hline
		Cough-time ($t$) & cough from each subject & 2, 5 to 100 with step of 5 \\
		
		\hline
		\hline
		\multicolumn{3}{c}{\textbf{\textit{Cough spotting}}}\\
		\hline
		\hline
		Frame length ($\mathcal{F}$) & used to extract features & $2^k$, $k=9, \ldots 12$ \\
		\hline
		No. of frames ($\mathcal{S}$) & extracted from audio & $10 \times k$, $k=7, 10, 12, 15$ \\
		
		\hline
		\hline
	\end{tabular}
\end{center}
\label{table:feat-hyper-parameter}
\end{table}


LR, LDA, SVM and MLP classifiers
were used to identify coughers and 
CNN, LSTM and Resnet50 were used to spot coughs as a trigger phrase.
Table \ref{table:class-hyper-parameter} lists the hyperparameters considered
and the ranges considered during the 5-fold \mbox{cross-validation}.
The standard deviation among the outer folds is noted as $\sigma_{ACC}$ in Table \ref{table:cougher-ident}. 
For Resnet50, the \mbox{50-layer} architecture described in 
\cite{he2016deep} 
has been used.

\begin{table}[h]
\footnotesize
\setlength{\tabcolsep}{4pt} 
\caption{\textbf{Classifier hyperparameters} used in both identifying `coughers' and spotting `cough'. } 
\centering 
\begin{center}
	\begin{tabular}{ c | c | c | c  }
		\hline
		\hline
		& \textbf{Hyperparameters} & \textbf{Classifier} & \textbf{Range} \\
		\hline
		\hline
		
		\multirow{5}{*}{\STAB{\rotatebox[origin=c]{90}{ coughers }}} & Regularisation & LR \& SVM & $10^i$ where $i=-7, \ldots 7$ \\
		\cline{2-4}
		& $l1$ penalty & LR  & 0 to 1 in steps of 0.05 \\
		\cline{2-4}
		& $l2$ penalty & LR, MLP  & 0 to 1 in steps of 0.05 \\
		\cline{2-4}
		& Kernel coeff. & SVM & $10^i$ where $i=-7, \ldots 7$  \\
		\cline{2-4}
		& No. of neurons & MLP & 70 to 150 in steps of 20 \\
		
		\hline
		
		\multirow{8}{*}{\STAB{\rotatebox[origin=c]{90}{cough }}} & Batch size & CNN \& LSTM & $2^k$ where $k=6, 7, 8$ \\
		\cline{2-4}
		& No. of epochs & CNN \& LSTM & 10 to 200 in steps of 20 \\
		\cline{2-4}
		& No. of conv filters & CNN & $3 \times 2^k$ where $k=3, 4, 5$ \\
		\cline{2-4}
		& kernel size & CNN & 2 and 3 \\
		\cline{2-4}
		& Dropout rate & CNN \& LSTM & 0.1 to 0.5 in steps of 0.2 \\
		\cline{2-4}
		& Dense layer size & CNN \& LSTM & $2^k$ where $k=4, 5$ \\
		\cline{2-4}
		& LSTM units & LSTM & $2^k$ where $k=6, 7, 8$ \\
		\cline{2-4}
		& Learning rate & LSTM & $10^k$ where $k=-2,-3,-4$ \\
		\hline
		\hline
	\end{tabular}
\end{center}
\label{table:class-hyper-parameter}
\end{table}


\section{Results and Discussion}



\begin{figure}
\centerline{\includegraphics[width=0.5\textwidth]{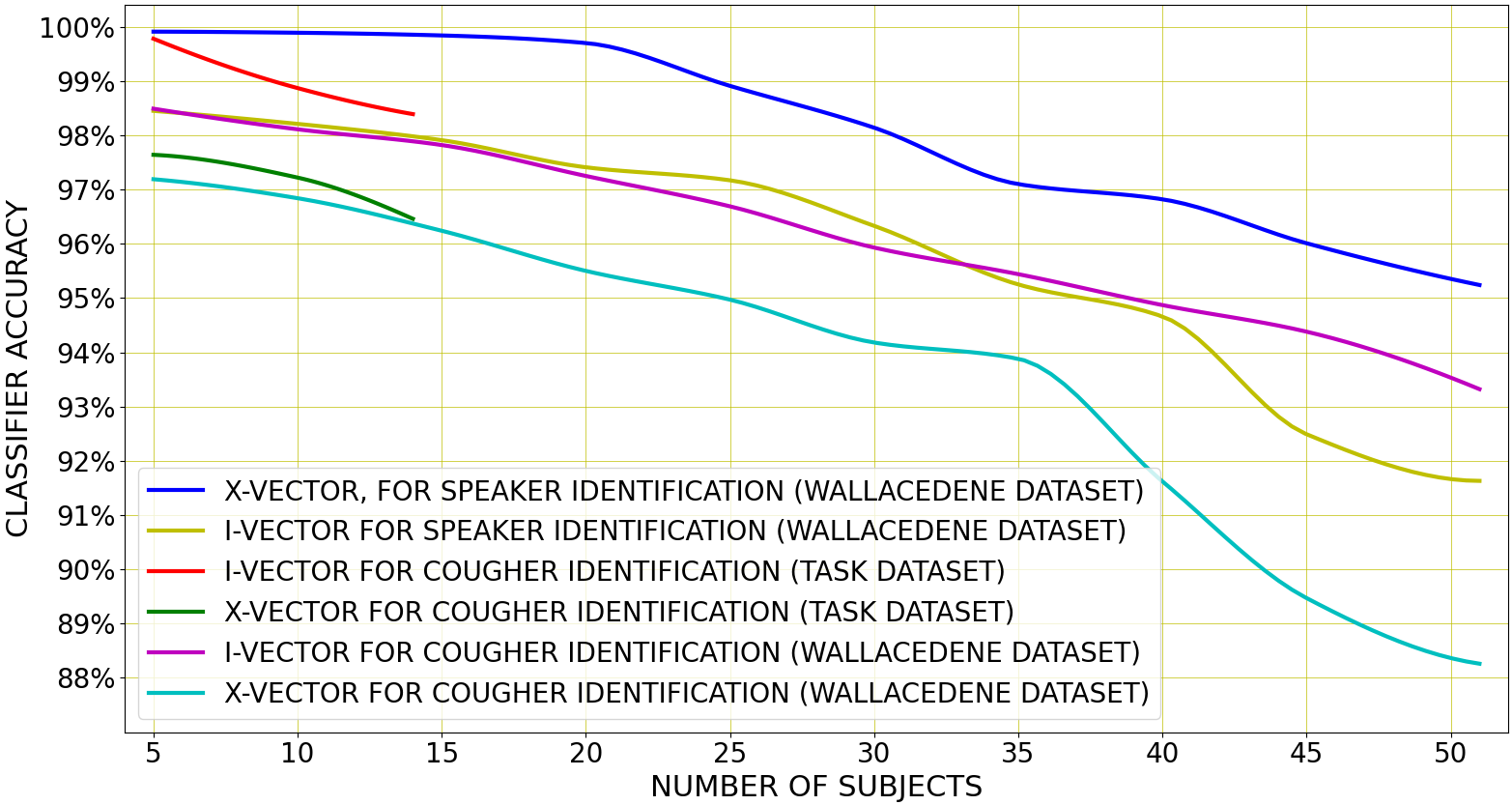}}
\caption{\textbf{Classifier performance.} The accuracies from the MLP classifier decrease while discriminating more subjects (Table \ref{table:cougher-ident}). }
\label{fig:coughers-acc}
\end{figure}

Table \ref{table:cougher-ident} shows the results using the best two features for both TASK (less-noisy) and Wallacedene (noisier) datasets. 
The highest accuracy (99.78\%) 
has been achieved by an MLP when using i-vectors to identify coughers from \mbox{100-sec} ($t=100$) long cough collected from each of 5 coughers. 
By increasing the number of coughers to 10 and 14, the performance of the MLP classifier decreased to 98.87\% and 98.39\% respectively for i-vectors (Table \ref{table:cougher-ident} and Figure \ref{fig:coughers-acc}). 
%
%

\begin{figure}
\centerline{\includegraphics[width=0.5\textwidth]{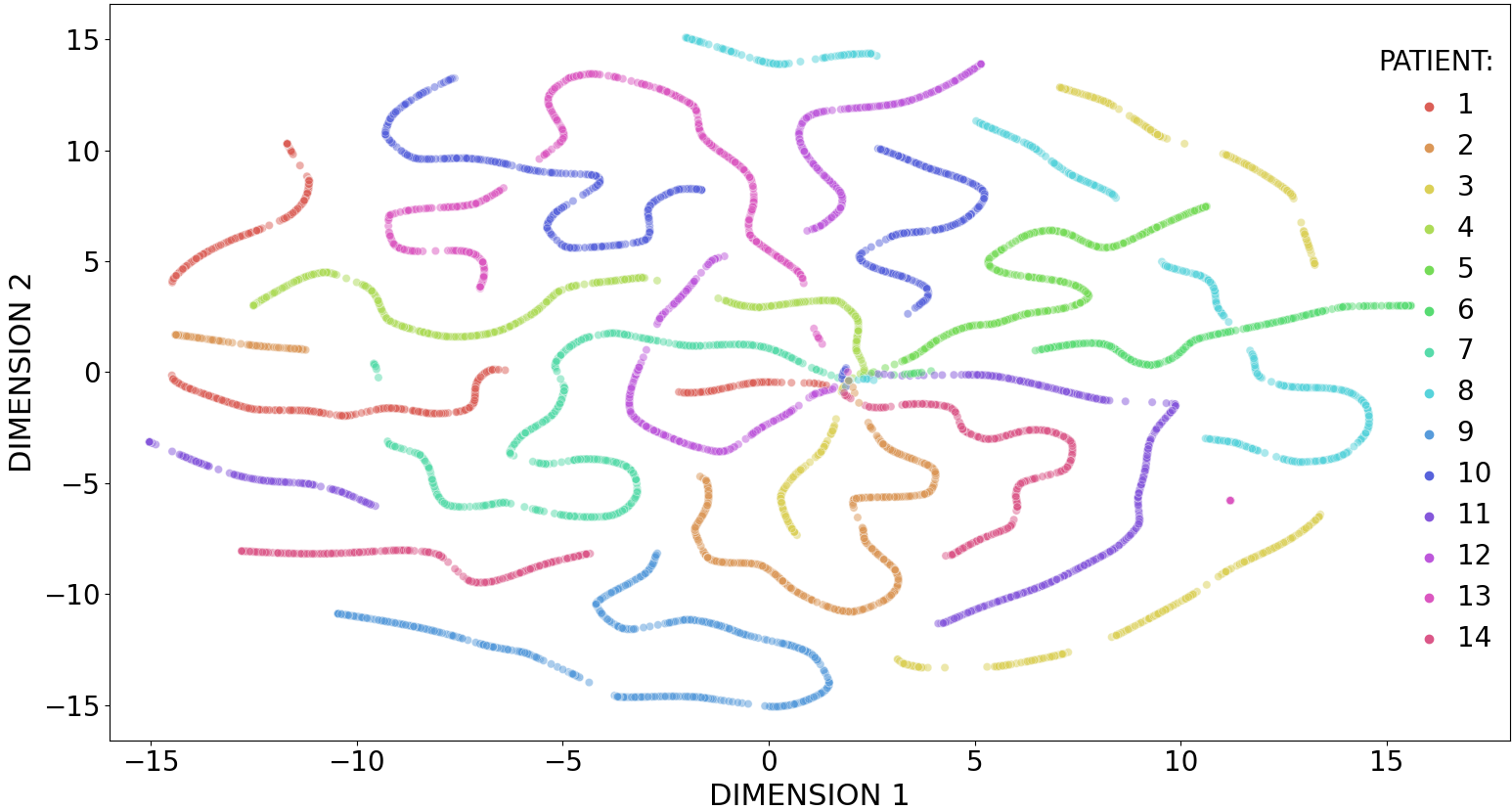}}
\caption{\textbf{The t-SNE cluster of i-vectors extracted from 2-sec long cough audio from 14 coughers in TASK dataset}. The MLP produces 95.11\% accuracy using these i-vectors in discriminating 14 coughers (Table \ref{table:cougher-ident}). }
\label{fig:tSNE}
\end{figure}
%

All classifiers performed well in identifying both coughers and speakers on the noisier the Wallacedene dataset. 
The speaker identification is used as the baseline and Table \ref{table:cougher-ident} shows that using x-vectors produced better classification scores than using i-vectors for speaker identification, as also found by others \cite{snyder2018x}. 
The highest accuracy (99.91\%) has been achieved using the MLP and x-vectors while discriminating among only 5 speakers. 
This accuracy drops to 98.14\% using MLP while differentiating between 30 speakers and to 95.24\% when discriminating among all 51 speakers in the Wallacedene dataset. 
For a smaller number of coughers, such as 5, the MLP classifier has achieved the highest accuracy of 98.49\% using i-vectors. 
As the number of coughers is increased to 15, 25, 40 and 51, 
the accuracy of the MLP has dropped to 97.82\%, 96.69\%, 94.87\% and 93.32\% respectively and the $\sigma_{ACC}$ has increased sharply. 
These scores show that although cougher identification is not as accurate as speaker identification, the performance is close, especially for a small number of subjects.

The results also show that, unsurprisingly, cougher identification on the less-noisy TASK dataset is more accurate than the noisier Wallacedene dataset. 
Although longer coughs from each subject improve the classifier accuracy in general, 
similar performance is achieved (accuracies of 95.11\% \& 90.02\% on the less-noisy \& the noisy data) for coughs as short as only 2 sec (Figure \ref{fig:tSNE}). 
Although the performance is close, i-vectors performed better than x-vectors in cougher identification.  
The MLP is the classifier of choice as it shows a lower $\sigma_{ACC}$ across the cross-validation folds for the \mbox{less-noisy} data than noisier data. 
d-vectors are outperformed by i-vectors and \mbox{x-vectors} for both speech and cough, as also found by \cite{li2015improved}, and thus excluded from Table \ref{table:cougher-ident}. 


\begin{table}[!h]
\scriptsize
\setlength{\tabcolsep}{4pt} 
\centering
\caption{\textbf{Classifier accuracies in identifying coughers} for both TASK and Wallacedene (WD) datasets.
}
\begin{tabular}{ c | c | c | c | c  c  c  c  c  }
	\hline
	\hline
	\textbf{Dataset} & \textbf{$N$} & $t$ & \textbf{Feature} & \textbf{LR} & \textbf{LDA} & \textbf{SVM} & \textbf{MLP} & \textbf{$\sigma_{ACC}$} \\
	\hline
	\hline
	
	{\multirow{5}{*}{TASK}} & {\multirow{2}{*}{5}} & 100 & i-vector & 98.91\% & 98.87\% & 99.44\% & \textit{99.78\%} & 0.0007 \\
	\cline{3-9}
	&  & 100 & x-vector & 96.71\% & 96.73\% & 97.54\% & 97.64\% & 0.0009 \\
	\cline{2-9}
	
	& {\multirow{2}{*}{10}} & 80 & i-vector & 97.54\% & 97.88\% & 98.19\% & \textit{98.87\%} & 0.0006 \\
	\cline{3-9}
	&  & 80 & x-vector & 96.31\% & 96.24\% & 96.55\% & 97.22\% & 0.0005 \\
	\cline{2-9}
	
	& {\multirow{3}{*}{14}} & 2 & i-vector & 94.41\% & 94.51\% & 94.55\% & \textbf{95.11\%} & 0.0005 \\
	\cline{3-9}
	&  & 100 & i-vector & 96.46\% & 96.71\% & 97.48\% & \textit{98.39\%} & 0.0006 \\
	\cline{3-9}
	&  & 100 & x-vector & 97.26\% & 97.54\% & 98.78\% & 96.46\% & 0.0008 \\
	\hline

	{\multirow{9}{*}{WD}} & {\multirow{2}{*}{5}} & 20 & i-vector & 97.23\% & 97.19\% & 97.77\% & \textit{98.49\%} & 0.0054 \\
	\cline{3-9}
	{\multirow{9}{*}{(Cougher)}} &  & 20 & x-vector & 95.54\% & 95.97\% & 96.72\% & 97.19\% & 0.0078 \\
	\cline{2-9}
	
	& {\multirow{2}{*}{15}} & 20 & i-vector & 97.16\% & 97.14\% & 97.31\% & \textit{97.82\%} & 0.0061 \\
	\cline{3-9}
	&  & 20 & x-vector & 95.41\% & 95.30\% & 95.72\% & 96.24\% & 0.0068 \\
	\cline{2-9}
	
	& {\multirow{2}{*}{25}} & 20 & i-vector & 95.04\% & 95.18\% & 95.94\% & \textit{96.69\%} & 0.0072 \\
	\cline{3-9}
	&  & 20 & x-vector & 93.31\% & 93.55\% & 94.07\% & 94.97\% & 0.0082 \\
	\cline{2-9}
	
	& {\multirow{2}{*}{40}} & 20 & i-vector & 93.38\% & 93.62\% & 94.09\% & \textit{94.87\%} & 0.0091 \\
	\cline{3-9}
	&  & 20 & x-vector & 90.23\% & 90.07\% & 90.97\% & 91.62\% & 0.0102 \\
	\cline{2-9}
	
	& {\multirow{3}{*}{51}} & 2 & i-vector & 89.26\% & 89.38\% & 89.22\% & \textbf{90.02\%} & 0.0178 \\
	\cline{3-9}
	&  & 20 & i-vector & 90.27\% & 90.49\% & 91.89\% & \textit{93.32\%} & 0.0301 \\
	\cline{3-9}
	&  & 20 & x-vector & 84.61\% & 84.74\% & 85.83\% & 88.26\% & 0.0247 \\
	\hline

	{\multirow{5}{*}{WD}} & {\multirow{2}{*}{5}} & --- & x-vector & 98.57\% & 98.64\% & 99.48\% & \textit{99.91\%} & 0.0018 \\
	\cline{3-9}
	{\multirow{5}{*}{(Speaker)}} &  & --- & i-vector & 97.21\% & 97.17\% & 97.70\% & 98.45\% & 0.0027 \\
	\cline{2-9}
	
	& {\multirow{2}{*}{30}} & --- & x-vector & 96.81\% & 96.85\% & 97.42\% & \textit{98.14\%} & 0.0081 \\
	\cline{3-9}
	&  & --- & i-vector & 94.81\% & 94.87\% & 95.18\% & 96.33\% & 0.0078 \\
	\cline{2-9}
	
	& {\multirow{2}{*}{51}} & --- & x-vector & 99.44\% & 99.44\% & 99.44\% & \textit{95.24\%} & 0.0229 \\
	\cline{3-9}
	&  & --- & i-vector & 90.01\% & 90.05\% & 90.34\% & 91.63\% & 0.0274 \\
	
	
	%
	
	\hline
	\hline
	
\end{tabular}
\label{table:cougher-ident}
\end{table}



\begin{figure}
\centerline{\includegraphics[width=0.5\textwidth]{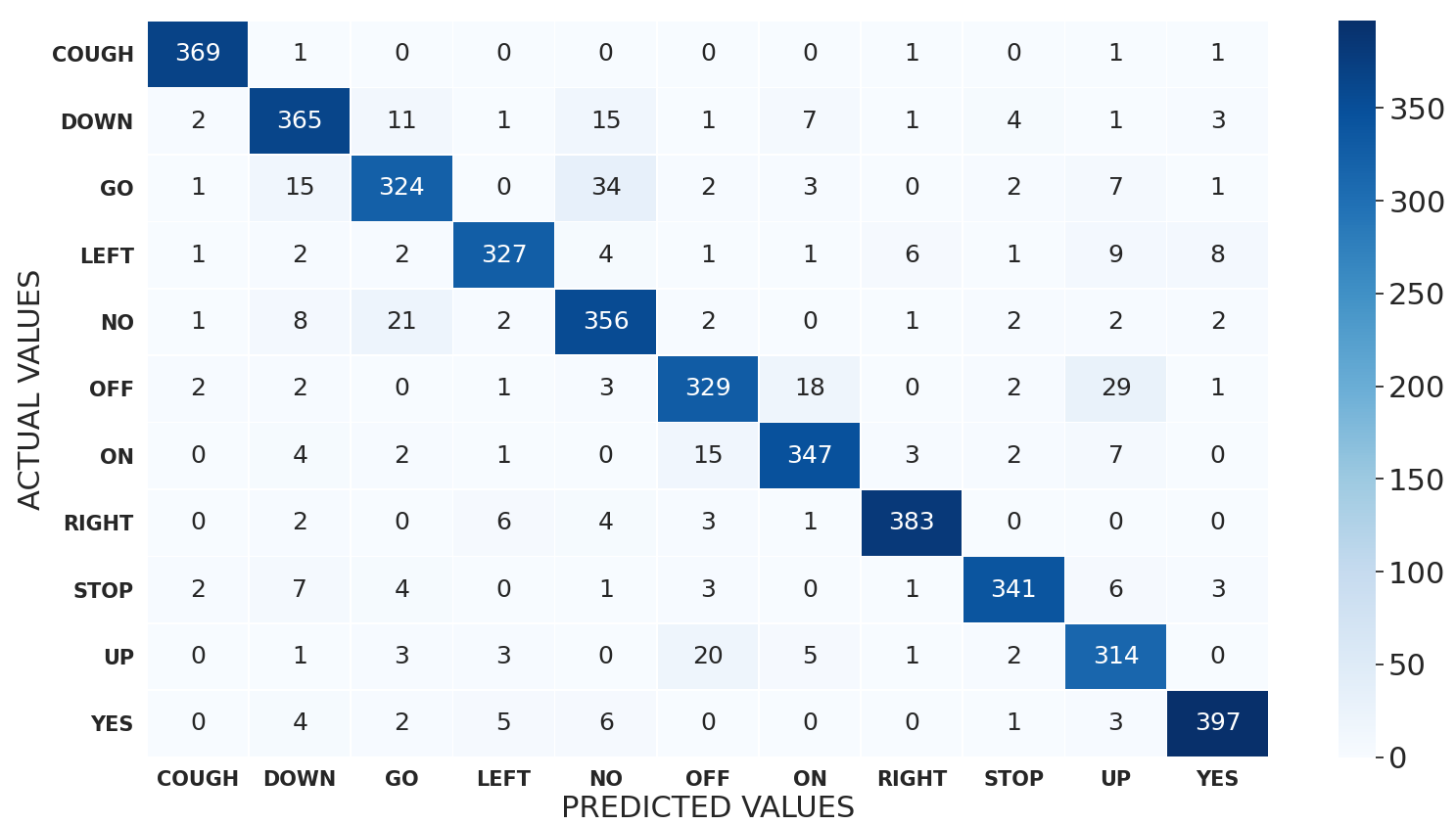}}
\caption{\textbf{The confusion matrix} of detecting coughs among 10 other trigger phrases in SC-11 dataset using the best Resnet50 classifier in Table \ref{table:cough-KWS}. 
}
\label{fig:conf-KWS}
\end{figure}

Coughs were successfully spotted among other trigger phrases in both the SC-11 and the SC-36 dataset. 
Table \ref{table:cough-KWS} shows that although LSTM and CNN have performed well, the best performance of 92.73\% accuracy ($\mathcal{ACC}$) \& mean Cohen's Kappa ($\mathcal{K}$) of 0.9218 on SC-11 and 88.58\% accuracy \& $\mathcal{K}$ of 0.8757 on SC-36 have been achieved using a Resnet50. 
The confusion matrix of the best SC-11 system exhibits an excellent performance for spotting coughs among the other trigger phrases in Figure \ref{fig:conf-KWS}. 
Table \ref{table:cough-KWS} also shows that the best CNN and Resnet50 results were obtained mostly when using 1024 sample (64 msec) long frames and 100 segments.


\begin{table}[!h]
\footnotesize
\setlength{\tabcolsep}{4pt} 
\centering
\caption{\textbf{Cough spotting:} The best-three results for each classifier shows Resnet50 has performed the best by achieving 92.73\% \& 88.58\% accuracy on the SC-11 \& SC-36 dataset. }
\begin{tabular}{ c | c | c | c | c | c | c | c | c }
	\hline
	\hline
	\multirow{2}{*}{\textbf{Classifier}} & \multicolumn{4}{c|}{\textbf{SC-11 Dataset}} & \multicolumn{4}{c}{\textbf{SC-36 Dataset}} \\
	
	\cline{2-9}
	
	& $\mathcal{F}$ & $\mathcal{S}$ & $\mathcal{ACC}$ & $\mathcal{K}$ & $\mathcal{F}$ & $\mathcal{S}$ & $\mathcal{ACC}$ & $\mathcal{K}$ \\
	
	\hline
	\hline
	\multirow{3}{*}{LSTM} & \textit{512} & \textit{150} & \textit{88.09\%} & \textit{0.8767} & \textit{512} & \textit{120} & \textit{80.74\%} & \textit{0.7937} \\
	\cline{2-9}
	& 2048 & 120 & 87.66\% & 0.8614 & 1024 & 120 & 80.40\% & 0.7931 \\
	\cline{2-9}
	& 512 & 70 & 87.09\% & 0.8598 & 512 & 100 & 80.11\% & 0.7902 \\ 		
	\hline
	
	\multirow{3}{*}{CNN} & \textit{1024} & \textit{100} & \textit{91.25\%} & \textit{0.9007} & \textit{1024} & \textit{120} & \textit{86.74\%} & \textit{0.8592} \\
	\cline{2-9}
	& 2048 & 100 & 90.72\% & 0.8981 & 1024 & 70 & 85.98\% & 0.8463 \\
	\cline{2-9}
	& 1024 & 70 & 90.11\% & 0.8945 & 2048 & 100 & 85.22\% & 0.8411 \\ 		
	\hline
	
	\multirow{3}{*}{Resnet50} & \textit{\textbf{1024}} & \textit{\textbf{100}} & \textit{\textbf{92.73\%}} & \textit{\textbf{0.9218}} & \textit{\textbf{2048}} & \textit{\textbf{100}} & \textit{\textbf{88.58\%}} & \textit{\textbf{0.8777}} \\
	\cline{2-9}
	& 2048 & 120 & 92.69\% & 0.8733 & 2048 & 70 & 87.94\% & 0.8729 \\
	\cline{2-9}
	& 2048 & 100 & 92.55\% & 0.8715 & 1024 & 120 & 87.68\% & 0.8702 \\ 		
	\hline
	\hline
	
\end{tabular}
\label{table:cough-KWS}
\end{table}


\section{Conclusion}
We propose a system using cough as a wake-word to spot coughs among other trigger phrases and identify the cougher.
A less-noisy and noisier dataset, containing 14 and 51 subjects respectively, were used to extract i-vectors, x-vectors and \mbox{d-vectors}, to classify the cougher.
The best performance was achieved using an MLP, showing coughers as many as 51 can be distinguished from one another with 90.02\% accuracy using i-vectors from as short as 2-sec long audio from each cougher
in the noisy environment. 
We also found that, unlike speakers, coughers were better identifiable using i-vectors.
Coughs were also spotted as wake-words using a Resnet50 on features keeping end-to-end \mbox{time-domain} information among 35 other keywords in the Google Speech Commands dataset with 88.58\% accuracy. 
Wake-cough represents a means of personalised, long-term cough monitoring system that is able to discriminate between coughers, non-intrusive and, due to the use of wake-word detection methods, power-efficient since a smartphone-based monitoring device can remain mostly dormant. 
Thus, it is an attractive and viable means for monitoring a patient's long-term recovery from lung ailments such as TB and COVID-19 in multi-bed ward environments.


In our future work, we aim to include more recent architectures 
and extend the dataset 
to investigate wake-cough's performance across age, gender etc. of the subjects and compare it with metric learning-based cougher identification \cite{jokic2022tripletcough}. 




%
%
%



\vspace{-5pt}

\bibliographystyle{IEEEbib}
\bibliography{reference}

\begin{thebibliography}{10}

\bibitem{schalkwyk2010your}
Johan Schalkwyk, Doug Beeferman, Fran{\c{c}}oise Beaufays, Bill Byrne, Ciprian
  Chelba, Mike Cohen, Maryam Kamvar, and Brian Strope,
\newblock ``{Y}our {W}ord is my {C}ommand'': {G}oogle search by {V}oice: {A}
  {C}ase {S}tudy,''
\newblock in {\em Advances in Speech Recognition}, pp. 61--90. Springer, 2010.

\bibitem{wu2018monophone}
Minhua Wu, Sankaran Panchapagesan, Ming Sun, Jiacheng Gu, Ryan Thomas, Shiv
  Naga~Prasad Vitaladevuni, Bjorn Hoffmeister, and Arindam Mandal,
\newblock ``Monophone-{B}ased {B}ackground {M}odeling for {T}wo-{S}tage
  {O}n-{D}evice {W}ake {W}ord {D}etection,''
\newblock in {\em 2018 IEEE International Conference on Acoustics, Speech and
  Signal Processing (ICASSP)}. IEEE, 2018, pp. 5494--5498.

\bibitem{gao2020towards}
Yixin Gao, Yuriy Mishchenko, Anish Shah, Spyros Matsoukas, and Shiv
  Vitaladevuni,
\newblock ``Towards {D}ata-{E}fficient {M}odeling for {W}ake {W}ord
  {S}potting,''
\newblock in {\em ICASSP 2020-2020 IEEE International Conference on Acoustics,
  Speech and Signal Processing (ICASSP)}. IEEE, 2020, pp. 7479--7483.

\bibitem{sainath2015convolutional}
Tara Sainath and Carolina Parada,
\newblock ``Convolutional {N}eural {N}etworks for {S}mall-footprint {K}eyword
  {S}potting,''
\newblock in {\em INTERSPEECH}, 2015, pp. 1478--1482.

\bibitem{chen2014small}
Guoguo Chen, Carolina Parada, and Georg Heigold,
\newblock ``Small-footprint keyword spotting using deep neural networks,''
\newblock in {\em 2014 IEEE International Conference on Acoustics, Speech and
  Signal Processing (ICASSP)}. IEEE, 2014, pp. 4087--4091.

\bibitem{botha2018detection}
Renier Botha, Grant Theron, Robbin Warren, Marisa Klopper, Keertan Dheda, Paul
  Van~Helden, and Thomas Niesler,
\newblock ``Detection of tuberculosis by automatic cough sound analysis,''
\newblock {\em Physiological Measurement}, vol. 39, no. 4, pp. 045005, 2018.

\bibitem{al2013signal}
Mahmood Al-khassaweneh and Ra’ed Bani~Abdelrahman,
\newblock ``A signal processing approach for the diagnosis of asthma from cough
  sounds,''
\newblock {\em Journal of Medical Engineering \& Technology}, vol. 37, no. 3,
  pp. 165--171, 2013.

\bibitem{pramono2016cough}
Renard Xaviero~Adhi Pramono, Syed~Anas Imtiaz, and Esther Rodriguez-Villegas,
\newblock ``A cough-based algorithm for automatic diagnosis of pertussis,''
\newblock {\em PLOS ONE}, vol. 11, no. 9, pp. e0162128, 2016.

\bibitem{pahar2020covid}
Madhurananda Pahar, Marisa Klopper, Robin Warren, and Thomas Niesler,
\newblock ``{COVID}-19 cough classification using machine learning and global
  smartphone recordings,''
\newblock {\em Computers in Biology and Medicine}, vol. 135, pp. 104572, 2021.

\bibitem{pahar2022covid}
Madhurananda Pahar, Marisa Klopper, Robin Warren, and Thomas Niesler,
\newblock ``{COVID}-19 detection in cough, breath and speech using deep
  transfer learning and bottleneck features,''
\newblock {\em Computers in Biology and Medicine}, vol. 141, pp. 105153, 2022.

\bibitem{ge2017deep}
Fengpei Ge and Yonghong Yan,
\newblock ``Deep neural network based wake-up-word speech recognition with
  two-stage detection,''
\newblock in {\em 2017 IEEE International Conference on Acoustics, Speech and
  Signal Processing (ICASSP)}. IEEE, 2017, pp. 2761--2765.

\bibitem{kepuska2009novel}
Veton~Z K{\"e}puska and TB~Klein,
\newblock ``A novel {W}ake-{U}p-{W}ord speech recognition system,
  {W}ake-{U}p-{W}ord recognition task, technology and evaluation,''
\newblock {\em Nonlinear Analysis: Theory, Methods \& Applications}, vol. 71,
  no. 12, pp. e2772--e2789, 2009.

\bibitem{senior2014improving}
Andrew Senior and Ignacio Lopez-Moreno,
\newblock ``Improving {DNN} speaker independence with i-vector inputs,''
\newblock in {\em 2014 IEEE International Conference on Acoustics, Speech and
  Signal Processing (ICASSP)}. IEEE, 2014, pp. 225--229.

\bibitem{snyder2018x}
David Snyder, Daniel Garcia-Romero, Gregory Sell, Daniel Povey, and Sanjeev
  Khudanpur,
\newblock ``X-{V}ectors: {R}obust {DNN} {E}mbeddings for {S}peaker
  {R}ecognition,''
\newblock in {\em 2018 IEEE International Conference on Acoustics, Speech and
  Signal Processing (ICASSP)}. IEEE, 2018, pp. 5329--5333.

\bibitem{wan2018generalized}
Li~Wan, Quan Wang, Alan Papir, and Ignacio~Lopez Moreno,
\newblock ``Generalized {E}nd-to-{E}nd {L}oss for {S}peaker {V}erification,''
\newblock in {\em 2018 IEEE International Conference on Acoustics, Speech and
  Signal Processing (ICASSP)}. IEEE, 2018, pp. 4879--4883.

\bibitem{whitehill2020whosecough}
Matt Whitehill, Jake Garrison, and Shwetak Patel,
\newblock ``Whosecough: {I}n-the-{W}ild {C}ougher {V}erification {U}sing
  {M}ultitask {L}earning,''
\newblock in {\em ICASSP 2020-2020 IEEE International Conference on Acoustics,
  Speech and Signal Processing (ICASSP)}. IEEE, 2020, pp. 896--900.

\bibitem{zhang2017speaker}
Miao Zhang, Yixiang Chen, Lantian Li, and Dong Wang,
\newblock ``Speaker recognition with cough, laugh and ``{W}ei",''
\newblock in {\em 2017 Asia-Pacific Signal and Information Processing
  Association Annual Summit and Conference (APSIPA ASC)}. IEEE, 2017, pp.
  497--501.

\bibitem{warden2018speech}
Pete Warden,
\newblock ``Speech {C}ommands: {A} {D}ataset for {L}imited-{V}ocabulary
  {S}peech {R}ecognition,''
\newblock {\em arXiv preprint arXiv:1804.03209}, 2018.

\bibitem{wittenburg2006elan}
Peter Wittenburg, Hennie Brugman, Albert Russel, Alex Klassmann, and Han
  Sloetjes,
\newblock ``{ELAN}: a professional framework for multimodality research,''
\newblock in {\em 5th International Conference on Language Resources and
  Evaluation (LREC 2006)}, 2006.

\bibitem{pahar2021deep}
Madhurananda Pahar, Igor Miranda, Andreas Diacon, and Thomas Niesler,
\newblock ``Deep {N}eural {N}etwork based {C}ough {D}etection using
  {B}ed-mounted {A}ccelerometer {M}easurements,''
\newblock in {\em ICASSP 2021 - 2021 IEEE International Conference on
  Acoustics, Speech and Signal Processing (ICASSP)}, 2021, pp. 8002--8006.

\bibitem{pahar2022automatic}
Madhurananda Pahar, Igor Miranda, Andreas Diacon, and Thomas Niesler,
\newblock ``Automatic {N}on-{I}nvasive {C}ough {D}etection based on
  {A}ccelerometer and {A}udio {S}ignals,''
\newblock {\em Journal of Signal Processing Systems}, pp. 1--15, 2022.

\bibitem{pahar2021tb}
Madhurananda Pahar, Marisa Klopper, Byron Reeve, Rob Warren, Grant Theron, and
  Thomas Niesler,
\newblock ``Automatic cough classification for tuberculosis screening in a
  real-world environment,''
\newblock {\em Physiological Measurement}, vol. 42, no. 10, pp. 105014, oct
  2021.

\bibitem{povey2011kaldi}
Daniel Povey, Arnab Ghoshal, Gilles Boulianne, Lukas Burget, Ondrej Glembek,
  Nagendra Goel, Mirko Hannemann, Petr Motlicek, Yanmin Qian, Petr Schwarz, Jan
  Silovsky, Georg Stemmer, and Karel Vesely,
\newblock ``The {K}aldi {S}peech {R}ecognition {T}oolkit,''
\newblock in {\em IEEE 2011 Workshop on Automatic Speech Recognition and
  Understanding}. IEEE Signal Processing Society, 2011,
\newblock IEEE Catalog No.: CFP11SRW-USB.

\bibitem{padhi2020}
Trideba Padhi, Astik Biswas, Febe de~Wet, Ewald van~der Westhuizen, and Thomas
  Niesler,
\newblock ``{Multilingual bottleneck features for improving ASR performance of
  code-switched speech in under-resourced languages},''
\newblock in {\em \textit{Proceedings of the First Workshop on Speech
  Technologies for Code-switching in Multilingual Communities} (WSTCSMC)},
  Shanghai, China, 2020.

\bibitem{he2016deep}
Kaiming He, Xiangyu Zhang, Shaoqing Ren, and Jian Sun,
\newblock ``Deep residual learning for image recognition,''
\newblock in {\em Proceedings of the IEEE conference on computer vision and
  pattern recognition}, 2016, pp. 770--778.

\bibitem{li2015improved}
Lantian Li, Yiye Lin, Zhiyong Zhang, and Dong Wang,
\newblock ``Improved deep speaker feature learning for text-dependent speaker
  recognition,''
\newblock in {\em 2015 Asia-Pacific Signal and Information Processing
  Association Annual Summit and Conference (APSIPA)}. IEEE, 2015, pp. 426--429.

\bibitem{jokic2022tripletcough}
Stefan Jokic, David Cleres, Frank Rassouli, Claudia Steurer-Stey, Milo~A.
  Puhan, Martin Brutsche, Elgar Fleisch, and Filipe Barata,
\newblock ``Triplet{C}ough: {C}ougher {I}dentification and {V}erification from
  {C}ontact-{F}ree {S}martphone-{B}ased {A}udio {R}ecordings {U}sing {M}etric
  {L}earning,''
\newblock {\em IEEE Journal of Biomedical and Health Informatics}, pp. 1--1,
  2022.

\end{thebibliography}

\end{document}